\documentclass[12pt]{article}
\usepackage[margin=1in]{geometry}
\usepackage[utf8]{inputenc}

\providecommand{\keywords}[1]
{
  \small	
  \textbf{\textit{Keywords}:} #1
}

\RequirePackage{longtable}
\RequirePackage{amsthm,amsmath,amsfonts,amssymb}
\RequirePackage[authoryear]{natbib}
\RequirePackage[colorlinks,citecolor=blue,urlcolor=blue]{hyperref}\RequirePackage{graphicx}\RequirePackage{comment}
\RequirePackage{bm}
\RequirePackage[capitalise]{cleveref}

\theoremstyle{definition}

\newcommand{\difflogit}[2]{\log{\frac{1+\exp{\left(-#1\right)}}{1+\exp{\left(-#2\right)}}} }

\newcommand{\bX}{\boldsymbol{X}}

\newcommand{\bSigma}{\boldsymbol{\Sigma}}
\newcommand{\bpsi}{\boldsymbol{\psi}}
\newcommand{\bbeta}{\boldsymbol{\beta}}
\newcommand{\btheta}{\boldsymbol{\theta}}
\newcommand{\set}[1]{\left\{#1\right\}}
\newcommand{\observedSample}{\set{\left(Y_{i}, \bX_{i}\right)}_{i=1}^{n}}
\newcommand{\pseudoSample}{\set{\left(Y_{i}^*, \bX_{i}^*\right)}_{i=1}^{M}}

\newcommand{\EE}[1]{\mathbb{E}\left\{#1\right\}}

\newcommand{\bmbeta}{\bm{\beta}}
\newcommand{\bmY}{\bm{Y}}

\newcommand{\bmX}{\bm{X}}
\newcommand{\bmvarepsilon}{\bm{\varepsilon}}
\newcommand{\calN}{\mathcal{N}}

\renewcommand{\log}[1]{\mathrm{log}\left(#1\right)}

\renewcommand{\exp}[1]{\mathrm{exp}\left\{#1\right\}}

\newcommand{\norm}[1]{\left\| #1 \right\|}

\newcommand{\diag}[1]{\mathrm{diag}\left\{#1\right\}}

\newcommand{\bbX}{\mathbb{X}}
\newcommand{\bbR}{\mathbb{R}}
\newcommand{\bs}[1]{\boldsymbol{#1}}
\newcommand{\vY}{{ \boldsymbol{\bm{Y}} }}
\newcommand{\mX}{{	\mathbb{X}			}}
\newcommand{\vX}{{ \boldsymbol{\bm{X}} }}

\newcommand{\tbmbeta}{\tilde{\bmbeta}}
\newcommand{\hbmbeta}{\hat{\bmbeta}}
\newcommand{\tp}{^{\top}}

\newcommand{\bD}{{ \boldsymbol{\bm{D}} }}
\renewcommand{\exp}{{\mbox{exp}}}

\renewcommand{\[}{\begin{equation}}
\renewcommand{\]}{\end{equation}}

\title{Catalytic Priors: Using Synthetic Data to Specify Prior Distributions in Bayesian Analysis}

\author{Dongming Huang\thanks{The first two authors contribute equally.}~\thanks{Department of Statistics and Data Science, National University of Singapore. Email: \texttt{stahd@nus.edu.sg}}, 
Feicheng Wang\footnotemark[1]~\thanks{Department of Statistics, Harvard University. Email: \texttt{fwangfeicheng@gmail.com}}, 
Donald B. Rubin\thanks{Department of Statistics, Harvard University. Email: \texttt{dbrubin@fas.harvard.edu}}, 
S. C. Kou\thanks{Department of Statistics and Department of Biostatistics, Harvard University. Email: \texttt{kou@stat.harvard.edu}}}

\begin{document}
\maketitle

\begin{abstract}Catalytic prior distributions provide general, easy-to-use, and interpretable specifications of prior distributions for Bayesian analysis. They are particularly beneficial when the observed data are inadequate to stably estimate a complex target model. 
A catalytic prior distribution is constructed by augmenting the observed data with synthetic data that are sampled from the predictive distribution of a simpler model estimated from the observed data. 
We illustrate the usefulness of the catalytic prior approach using an example from labor economics.  
In the example, 
the resulting Bayesian inference reflects many important aspects of the observed data, 
and the estimation accuracy and predictive performance of the inference based on the catalytic prior are superior to, or comparable to, that of other commonly used prior distributions. 
We further explore the connection between the catalytic prior approach and a few popular regularization methods. We expect the catalytic prior approach to be useful in many applications. 
\newline \keywords{
{synthetic data},
{insufficient data},
{stable estimation},
{regularization methods}
}
\end{abstract}

\section{Introduction}\label{sec:intro}

The specification of the prior distribution can be one of the most important yet challenging tasks in Bayesian data analysis.
Essentially every Bayesian analysis is based on the posterior distribution, which combines the prior distribution with the data, as represented by the likelihood function. 
With abundant observed data, the information from the data dominates the contribution of the prior distribution in the sense that the choice of prior distribution has little effect on the posterior distribution. When data are scarce, however, the prior distribution plays an important role in determining the posterior distribution and should be chosen carefully. 
Ideally, the prior should 
1) be simple and interpretable, 
2) provide stable posterior inference even when sample sizes are small, 
3) quantify existing knowledge about the scientific problem, and
4) be easy to implement. 
This paper discusses a universal method to construct such priors, which were called \textit{catalytic priors} in \cite{huang2020catalytic}.

The central idea behind the catalytic prior construction is from the viewpoint that data are real whereas models are imperfect mathematical constructions used by researchers to understand data. This perspective accords with the celebrated aphorism from George Box that
\textit{``all models are wrong, but some are useful"} \citep{BOX1979201} and the earlier comment from John von Neumann that \textit{``truth ... is much too complicated to allow anything but approximations"} \citep{von1947mathematician}.
From this data-centric viewpoint, a prior distribution amounts to ``prior data''. 
For instance, to infer the probability of a Bernoulli trial, the beta prior, $\text{Beta}(\alpha, \beta)$, is equivalent to 
supplementing the actual data with $\alpha$ successes and $\beta$ failures. 
For related statistical discussion dating back to at least the 1960s, see \citet{birnbaum1962foundations}, \citet{savage1962foundations}, and \citet{pratt1964foundations,pratt1965bayesian,dempster1968generalization,leamer1976bayesian,clogg1991multiple}.

For a general statistical model, the construction of a catalytic prior proceeds by generating synthetic prior data from a simpler model and then using these synthetic data to represent the prior distribution. The method's name, \textit{catalytic prior}, is inspired by the concept of catalysts in chemistry: a catalyst stimulates a reaction to take place; without the catalyst, the reaction cannot take place efficiently, but only a small amount of the catalyst is needed. In this analogy, the synthetic data play a supporting role akin to the catalyst, without which it can be difficult to make a stable inference.

\begin{figure*}
\caption{Process of the Catalytic Prior Construction. }
\centering
\includegraphics[width=0.7\textwidth]{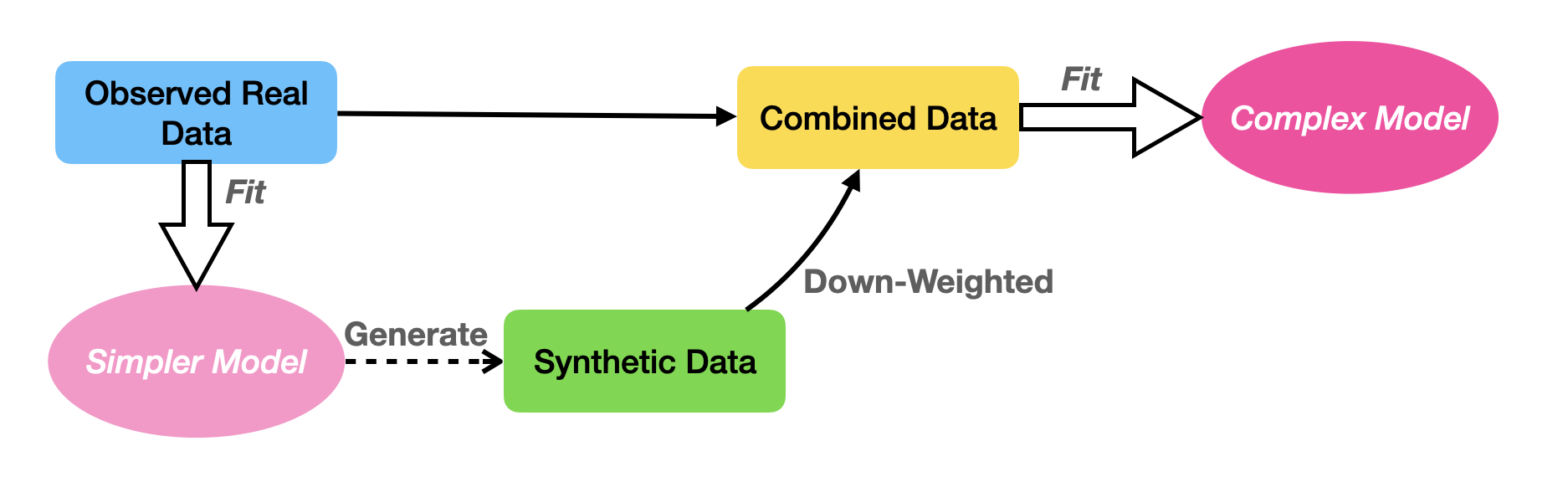}
\label{fig:construct}
\end{figure*}

Inference based on a catalytic prior is operationally described by the following procedure. 
Suppose we have a complex target model that we want to fit to the observed data $(\mX,\vY)$, where $\mX$ represents observed predictor variables and $
\vY$ represents observed response/outcome data, but we do not have enough observations to fit this complex model. We first fit a simpler model to the observed data $(\mX,\vY)$ that is easy to implement using standard software, and then generate synthetic data $(\mX^*, \vY^*)$ from this simpler model. 
Next we combine the synthetic data with the actual observed data, where the synthetic data receive less weight than the observed data because they are not actually observed. 
We then fit the complex model (i.e., our target that we want to analyze) to the combined data (i.e., the observed data combined with the down-weighted synthetic data); 
statistical inferences and decisions regarding the complex target model are then made using standard software applied to the combined data.
This process is illustrated in \cref{fig:construct}. 

When a catalytic prior is used, inferences about the complex target model are ``pulled'' towards the simpler model. 
In this way, a stable inference can be made even when the observed data are insufficient to fit the complex model using the same standard procedures. 

To illustrate a catalytic prior, consider the example of high-dimensional linear regression $\bmY = \bbX \bbeta + \bmvarepsilon$, 
where $\bmvarepsilon$ is a normal random vector and $\bbeta$ is the vector of unknown parameters
of dimension $p$, larger than 
the sample size $n$. 
There are infinitely many points that maximize the likelihood function, so maximum likelihood estimation of $\bbeta$ is not unique. 
To address this problem, we generate synthetic covariates $\bbX^*$ and synthetic responses $\bmY^*$, but down-weight them before  combining them with the observed data.
Standard analysis is then implemented on the combined data. For example, a unique maximum likelihood estimate of $\bbeta$ based on the combined ``data'' exists if the synthetic covariate matrix $\bbX^*$ has rank $p$. 
In \cref{sec:def}, we formally define the catalytic prior and continue this example. Using this catalytic prior distribution, one can easily formulate the resulting posterior distribution and compute the resulting posterior mode (the maximum a posteriori estimate) using standard statistical software.

These catalytic priors have simple interpretations and are easy to formulate.
\citet{huang2020catalytic} studied catalytic prior in the context of generalized linear models and discussed its possible advantages. 
The current paper extends the idea and its discussion by demonstrating the use of catalytic priors in an example taken from labor economics, offering practical guidelines and further exploring the connection between the catalytic prior approach and classical regularization methods.

\section{Catalytic priors} \label{sec:def}

\subsection{Construction}
Let $\observedSample$ be $n$ independent pairs of observed data, where $Y_i$ is a response and $\bX_i$ is a vector of covariates or predictors. 
Suppose the goal is to use $\observedSample$ to study a complex (e.g., high-dimensional) target model for $Y_i$ given $\bX_{i}$ and $\btheta$, where $\btheta$ is the unknown target parameter of primary interest, possibly of higher dimension than the sample size $n$:
\begin{equation}
    \label{eq: complex model}
    Y_{i} \mid \bX_{i}, \btheta \sim f\left(y \mid \bX_{i}, \btheta\right), i=1,2, \ldots, n.
\end{equation}

A catalytic prior generates synthetic data from a simpler model $g$, and uses the synthetic data as a prior sample for fitting the complex target model $f$. 
More concretely, the strategy is to generate $M$ synthetic data points  $\pseudoSample$ from a simpler model: 
\begin{equation}
    \label{eq: simple model}
    \bX_{i}^{*}\sim Q(x),\quad Y_{i}^{*}\mid \bX_{i}^{*} \sim g(y|\bX_{i}^*, \bpsi), 
\end{equation}
where $g$ is the simpler model with the parameter $\bpsi$ estimated from the observed data $\observedSample$, and $Q$ denotes the distribution from which the synthetic covariates $\bX^{*}$ are generated.
Because the synthetic data are not actually observed, 
they are down-weighted using a positive tuning parameter $\tau$ that determines their total weight
--- intuitively, the total weight of the synthetic data corresponds to $\tau$ data points, each assigned a weight of $\tau/M$. 
The catalytic prior is then defined to be
\begin{equation}
\label{eq: catalytic prior finite sample}
\begin{aligned}
\pi_{cat,M}(\btheta \mid \tau) \propto & \left\{\prod_{i=1}^{M} f\left(Y_{i}^{*} \mid \boldsymbol{X}_{i}^{*}, \btheta\right)\right\}^{\tau / M} \\
= & \exp{
\left\{ \frac{\tau}{M}
\sum_{i=1}^M \log{
f\left(Y_{i}^{*} \mid \bX_{i}^{*},  \btheta\right)}
\right\} 
},
\end{aligned}
\end{equation}
where $f$ is the density function in the complex target model as given in \cref{eq: complex model}.
When $M$ goes to infinity, we have the corresponding population form of this catalytic prior distribution for $\theta$ with weight $\tau$:
\begin{equation}
\pi_{cat,\infty}(\btheta \mid \tau) \propto
\exp{
\left\{ \tau
\mathbb{E}_{Y^*, \bX^*} \log{
f\left(Y_{}^{*} \mid \bX_{}^{*},  \btheta\right)}
\right\}
}. 
\label{eq: pop cata}
\end{equation}
Here the expectation is taken with respect to $Y^{*} \sim g(\cdot|\bX_{}^{*})$ and $\bX_{}^{*} \sim  Q(\cdot )$, as in \cref{eq: simple model}.
Because the catalytic prior has the same form as the likelihood based on the actual data, the posterior density can be immediately obtained: 
\begin{equation}
\label{eq:cata_posterior}
\begin{aligned}
\pi(\btheta \mid \mathbb{X}, \boldsymbol{Y}, \tau) \propto & ~ \pi_{cat,M}(\btheta \mid \tau) f(\boldsymbol{Y} \mid \mathbb{X}, \btheta) \\
\propto & ~ \exp \left(\frac { \tau } { M } \sum _ { i = 1 } ^ { M } \operatorname { l o g } \left(f\left(Y_{i}^{*} \mid \boldsymbol{X}_{i}^{*}, \btheta\right)\right) \right. \\ 
& ~ \left. +\sum_{i=1}^{n} \log{f\left(Y_{i} \mid \boldsymbol{X}_{i}, \btheta \right) } \right). 
\end{aligned} 
\end{equation}

\subsection{Illustration in linear regression}\label{sec:example linear regression}
To illustrate, consider the simple standard linear regression model $\bmY = \bbX \bmbeta + \bmvarepsilon$, where the residuals $\bmvarepsilon$ are i.i.d. Gaussian with mean 0 and known standard deviation $\sigma$, and 
$\bbeta$ is the unknown $p$-dimensional vector of regression coefficients. 
Maximum likelihood estimation of $\bmbeta$ encounters difficulty when $p$ is larger than the sample size $n$. 
To construct the catalytic prior, we can choose the simpler model $g$ to be a sub-model of $f$ that only involves a pre-selected set of components of $\bX$; we call this set $S$. We then fit the simpler model to the observed data $(\bbX, \bmY)$, and denote by $\tbmbeta_{S}$ the fitted parameter, which has zero coefficients for components of $\bX$ not in $S$. 
We then generate a synthetic covariate matrix $\bbX^*$ of dimension $M\times p$ that has rank $p$, where, as above, $M$ is the number of synthetic data points. We then generate $M$ synthetic responses from the fitted model: $\bmY^* \mid  \bbX^*  \sim \calN( \bbX^*\tbmbeta_S, \sigma^2 \mathbb{I}_{M} )$.

According to \cref{eq: catalytic prior finite sample}, the catalytic prior can then be expressed as:
$$
\begin{aligned}
\pi_{cat,M}(\bmbeta \mid \tau) \propto
\exp\left\{\frac{\tau}{M} 
\left( 
    - \frac{1}{2\sigma^2} \| \bmY^*- \bbX^*
    \bmbeta\|^2 
\right)
\right\}.
\end{aligned} 
$$
Here $\tau$ is the total weighting parameter. % with fixed $M$. 
The larger $\tau$, the larger impact the synthetic data will have on the resulting posterior distribution of $\bbeta$. 
The number of synthetic data points, $M$, should be at least as large as $p$ to ensure stable results. Simple algebra shows that, in this example, the catalytic prior reduces to the normal prior
$$
\bmbeta \sim \calN\left(\widehat{\bmbeta}_0, \frac{\sigma^2}{\tau}
\left(\frac{1}{M}(\bbX^*)^\top \bbX^*\right)^{-1}
\right),
$$
where $\widehat{\bmbeta}_0=\left( (\bbX^*)^\top \bbX^*\right)^{-1} (\bbX^*)^\top \bmY^* $ is the maximum likelihood estimate of the complex target model based only on the synthetic data. 
The posterior distribution is proportional to the product of the prior and the likelihood, which here is also a normal distribution with mean 
$$
\hbmbeta = \left(\bbX^\top \bbX + \frac{\tau}{M}(\bbX^*)^\top \bbX^*\right)^{-1}
\left(
\bbX^\top \bmY + \frac{\tau}{M}
(\bbX^*)^\top \bmY^*
\right)
$$
and covariance matrix 
$$
\widehat\Sigma = \sigma^2\left(\bbX^\top \bbX + \frac{\tau}{M}(\bbX^*)^\top \bbX^*\right)^{-1}.
$$

\subsection{Specification and properties}\label{sec:cata spec}
We now briefly discuss how to specify the components of the catalytic prior in general. 

Examples of a simpler model $g$ include the constant (i.e., intercept-only) model, %\citep{clogg1991multiple}, 
a model with only a few important predictor variables, and models nested in the complex target model. 
We want to emphasize that the simpler model $g$ does not have to be a sub-model of the complex target model $f$; for example, $f$ could be a tree model, where $g$ could be a linear model with a handful of predictors.  
When there are multiple suitable candidates for the simpler model, we can generate synthetic responses based on a mixture of the simpler models (a mixture of the fitted predictive distributions, to be precise);  see \cref{sec: more sources of synthetic data} for an example.
When there is no or little prior knowledge available, we suggest using a very simple model for $g$.

Synthetic covariates can be drawn from a fixed distribution; for example, $\set{\bX_i^*}_{i=1}^M$ can be generated by direct resampling from the observed $\set{\bX_i}_{i=1}^n$ or sampling from each individual covariate's marginal distribution to generate covariates that are mutually independent. 
A detailed discussion on the generation of synthetic covariates can be found in Supplementary Information of  \cite{huang2020catalytic}.

The synthetic sample size $M$ should be as large as possible given the time and computational resources. 
However, increasing $M$ beyond a certain point will yield diminishing returns, because the variability of synthetic data would eventually have negligible impact on the prior and the posterior. 
Based on our experience, if the dimension $p$ of the target model is no more than $20$, setting $M$ as small as 400 usually suffices to guarantee satisfactory posterior inference. 
If the target model is a generalized linear model, the theory presented in \cite{huang2020catalytic} suggests using $M$ greater than $4p$ is sufficient to ensure the properness of the catalytic prior. 
As there does not appear to be a universal value of $M$ that suits all problems, we advise the users to experiment with different values of $M$, taking into consideration the scale of their specific problem and available computational resources.

Specifying the total weight $\tau$ is important, as this parameter governs the degree of shrinkage toward the simpler model.
A simple but useful choice is $\tau=p$, which means that the effective number of prior data-points is the same as the dimension of the complex target model.

Another choice is to let $\tau$ grow with $n$ but at a slower rate. This choice allows the total weight to depend on the sample size $n$ while still controlling the impact of synthetic data. 
In general, the shrinkage toward the simpler model is controlled by choosing a suitable value of $\tau$. For example, if a user suspects that the true values of the parameters might be large, a small $\tau$ can be specified to minimize excessive shrinkage. 
Two other systematic ways to specify $\tau$ are by tuning the parameter through frequentist predictive risk estimation, and by putting a Gamma hyperprior on $\tau$, both discussed in \cite{huang2020catalytic}.

One appealing aspect of the catalytic prior is its general applicability. The expression of a catalytic prior in \cref{eq: catalytic prior finite sample} is as simple as the weighted likelihood of the synthetic data, and the application of a catalytic prior is computationally no more difficult than appending $M$ weighted observations to the observed data (see \cref{eq:cata_posterior}). 
For example, to implement  posterior inference with a catalytic prior, one can first generate synthetic data and merge them with the observed data, and then include a weight vector (weight $1$ for each observed data point and weight $\tau/M$ for each synthetic data point) to standard model fitting algorithms (such as the regression functions \texttt{lm()} or \texttt{glm()} in the \textbf{R} programming language). 
Apart from generating synthetic data, there is little extra programming burden. 
The use of synthetic data makes it straightforward to interpret what information is provided by a catalytic prior. 

Because the construction of a catalytic prior only depends on the weighted likelihood, it is invariant to any invertible affine transformation of the parameters of the complex model.
This is an attractive property for regression analysis when categorical variables are coded as quantitative variables (for example, through ``dummy'' variables) or when numerical variables may be shifted and scaled; 
if two analysts use different coding systems for the categorical variables and use different centers and scaling factors for the numerical variables, they will obtain equivalent catalytic priors (as long as there is a 1-1 correspondence between the two coding systems). 
From this perspective, catalytic priors may be preferable to other priors that do not possess this property, such as the class of multivariate normal priors. 

\section{An example from labor economics}\label{sec:app-swim}
We illustrate the application of catalytic priors in a real-world example from labor economics. 
In this example, to delineate the strengths and weaknesses of different methods, we now compare the performance of the result from a catalytic prior to that from the flat prior as well as the result from a popular prior, the Cauchy prior, which was proposed specifically for logistic regression models by \citet{gelman2008weakly}.

\subsection{The SWIM program and the models}

The Saturation Work Initiative Model (SWIM) in San Diego was a social program from 1985 to 1987 designed to promote labor market outcomes for individuals who were eligible for the Aid to Families with Dependent Children Program \citep{friedlander1993saturation}; also see \citet{friedlander1992high,friedlander1995evaluating,hotz2005predicting,imbens2015causal}.
SWIM provided job search assistance, unpaid work experience, education, and training for its participants.
The enrollees in the study were randomly assigned to either participate in SWIM or not. 
In the original study, the outcomes of interest are post-program employment status (whether employed during the follow-up period after the program) 
and earnings (if employed).  
For the program participants, the pre-program covariates available to us include demographics and recent labor market histories. 
In our illustrative analysis, we use the same subset of covariates as in \citet[Chapter 11]{imbens2015causal}, which are: 
\begin{itemize}
\item Indicator variables: (1) gender; (2) age over 35; (3) having a high school diploma; (4) never married; (5) divorced or widowed; (6) having children under 6; (7) racially identified as African American; (8) ethnically identified as Hispanic; (9) having positive earnings in the previous year;
\item Numeric variables: (10) number of children; and (11) earnings in the previous year.
\end{itemize}

In our analysis, we focus on the effect of SWIM on the post-program employment status $Y$, where $Y=1$ means employed within one year after the program and $Y=0$ otherwise. 
For enrollee $i$, let $Z_i$ denote 
the indicator of the (randomized) program participation ($Z_i=1$ means assigned to participate in the program and $Z_i=0$ otherwise). Let $Y_i(z)$ denote the post-program employment status given assignment $z$ ($z=0,1$); $Y_i(1)$ and $Y_i(0)$ are referred to as the ``potential outcomes" because it is impossible to observe them both: only the one corresponding to $z=Z_i$ can be observed. Under the framework of the Rubin Causal Model \citep{rubin1974estimating,holland1986statistics}, the causal effect for an individual compares the potential outcomes of this individual under $z=1$ and $z=0$, i.e., the contrast between $Y_i(1)$ and $Y_i(0)$. Note that we have implicitly assumed that potential outcomes for enrollee $i$ are unaffected by the treatment assignments of other enrollees, which, when combined with the no hidden version of treatment assumption, is known as the \textit{Stable Unit Treatment Value Assumption} (SUTVA) \citep{rubin1980randomization}. 

 For enrollee $i$, we assume conditional independence between the potential outcomes given covariates, and model them by two independent logistic regression models
\begin{align}
   \label{eq:SWIM-model-Z1}
    \text{logit} ~ \mathbb{P} \left( Y_{i}(1)=1 \mid \bmX_{i}, \bmbeta_{t}\right) & =  \bmX_{i}\tp \bmbeta_{t},  \\ 
   \label{eq:SWIM-model-Z0}
    \text{logit} ~ \mathbb{P} \left( Y_{i}(0)=1 \mid \bmX_{i}, \bmbeta_{c} \right) & =  \bmX_{i}\tp \bmbeta_{c}, 
\end{align}
where $\text{logit}(q)=\log{q/(1-q)}$ and $\bmX_{i}$ is the vector of covariates (including the constant term $1$ for the intercept). 
We assume conditional independence here to demonstrate the main idea of our method. One can apply our method to a more general formulation that allows for conditional dependence between a pair of potential outcomes.
The unit-level causal effect of the program participation on the post-program employment status for enrollee $i$ can be measured by the log probability ratio 
\begin{equation}\label{eq:swim-log-prob-ratio}
\begin{aligned}
\gamma_i := & \log{ \frac{\mathbb{P}(Y_{i}(1)=1\mid  \bmX_{i}, \bmbeta_{t}) } {\mathbb{P}( Y_{i}(0)=1\mid  \bmX_{i}, \bmbeta_{c}) }} \\
= & \difflogit{\bmX_{i} \tp \bmbeta_{c}}{\bmX_{i} \tp \bmbeta_{t}}. 
\end{aligned}
\end{equation}
$\gamma_i=0$ means that for enrollee $i$, the program participation does not affect the probability of employment. If $\gamma_i>0$, participating in the program is beneficial for enrollee $i$ because it increases the prospect of employment. The average value of $\gamma_i$, namely,
\begin{equation}\label{eq:avg_gamma}
\gamma_{avg} = \mbox{average of }\gamma_i \mbox{ over all enrollees} 
\end{equation}
measures the overall effect of the program participation.

To formulate the catalytic prior, we first generate synthetic data $\{(Y_{i}^{*}(1),Y_{i}^{*}(0), \bmX_{i}^{*})\}_{i=1}^{M}$. It is interesting to note that unlike the actual observations, where either $Y(1)$ or $Y(0)$ is unobserved, the synthetic data perspective allows us to generate both $Y^{*}(1)$ and $Y^{*}(0)$.
Under the independence assumption of the potential outcomes for each enrollee (\cref{eq:SWIM-model-Z1,eq:SWIM-model-Z0}), the catalytic prior for the coefficients $(\bmbeta_{t},\bmbeta_{c})$ is the (independent) product $$\pi_{t,M}(\bmbeta_{t}\mid \tau)\pi_{c,M}(\bmbeta_{c}\mid \tau),$$
where 
\begin{equation*}
           \pi_{t,M}(\bmbeta_{t}\mid \tau)
        \propto
        \left\{\prod_{i=1}^{M} \frac{\exp\left\{Y_{i}^{*}(1)(\boldsymbol{X}_{i}^{*})^\top\bmbeta_{t}\right\}}
        {1 + \exp\left\{(\boldsymbol{X}_{i}^{*})^\top\bmbeta_{t}\right\}}\right\}^{\tau / M} 
\end{equation*}
and $\pi_{c,M}(\bmbeta_{c}\mid \tau)$ is expressed analogously with $Y_{i}^{*}(1)$ replaced by $Y_{i}^{*}(0)$ and $\bmbeta_{t}$ replaced by $\bmbeta_{c}$. 

By the complete randomization design of the study, the assignment indicators $Z_i$'s are independent of the enrollees' potential outcomes and the covariates. Thus, 
the conditional distribution of the observed outcomes given the parameters is 
\begin{equation}
\label{eq:swim-likelihood}
\begin{aligned}
& \mathbb{P}\left(  \{ Y_i^{obs} \}_{1}^n \mid \bmbeta_{t}, \bmbeta_{c}, \{Z_i\}_{1}^n,  \{\bmX_i\}_{1}^n \right) \\
= & \prod_{i: Z_i=1}f\left(Y_i^{obs} , \bmX_{i}\tp \bmbeta_{t}\right)\prod_{i: Z_i=0}f\left(Y_i^{obs} , \bmX_{i}\tp \bmbeta_{c}\right),
\end{aligned}
\end{equation}
where $Y_{i}^{obs}=Y_i(Z_i)$ represents the observed response and $f(a,b)=e^{ab}/(1+e^{b})$; see \citet[Section 8.4.2]{imbens2015causal} . 
Because $\bmbeta_{t}$ and $\bmbeta_{c}$ are independent under this catalytic prior, 
\cref{eq:swim-likelihood} allows us to find the posterior distributions for $\bmbeta_{t}$ and $\bmbeta_{c}$ separately by fitting logistic regression models to the treatment group and the control group respectively. 

In our implementation of the catalytic prior, we set the total weight parameter $\tau$ equal to the total number of parameters to be estimated, i.e., $\tau=$ dimension of $\bmbeta_{t} +$  dimension of $\bmbeta_{c}$ $=12+12=24$. For the synthetic sample size $M$, we choose a value of $400$, a large number relative to the dimension of model parameters. 
To generate a synthetic data point $(Y_{i}^{*}(1),Y_{i}^{*}(0), \bmX_{i}^{*})$, each synthetic covariate variable in $\bmX_{i}^{*}$ is component-wise independently resampled from the actual observations. The synthetic response $Y^*(1)$ is set to be $\hat{\mu}_t$, which is the proportion of the observed $Y(1) = 1$ in the treatment group\footnote{Alternatively, we could draw $Y^*(1)$ from $\text{Bernoulli}(\hat{\mu}_t)$ but taking $Y^*(1)$ to be the expected value of the Bernoulli distribution is particularly simple.}. The synthetic response $Y^*(0)$ is set to be $\hat{\mu}_c$, the proportion of the observed $Y(0) = 1$ in the control group. 

\subsection{Analyzing the full data set with the catalytic prior}
\begin{figure*}%[!h]
\centering
\includegraphics[width=\textwidth]{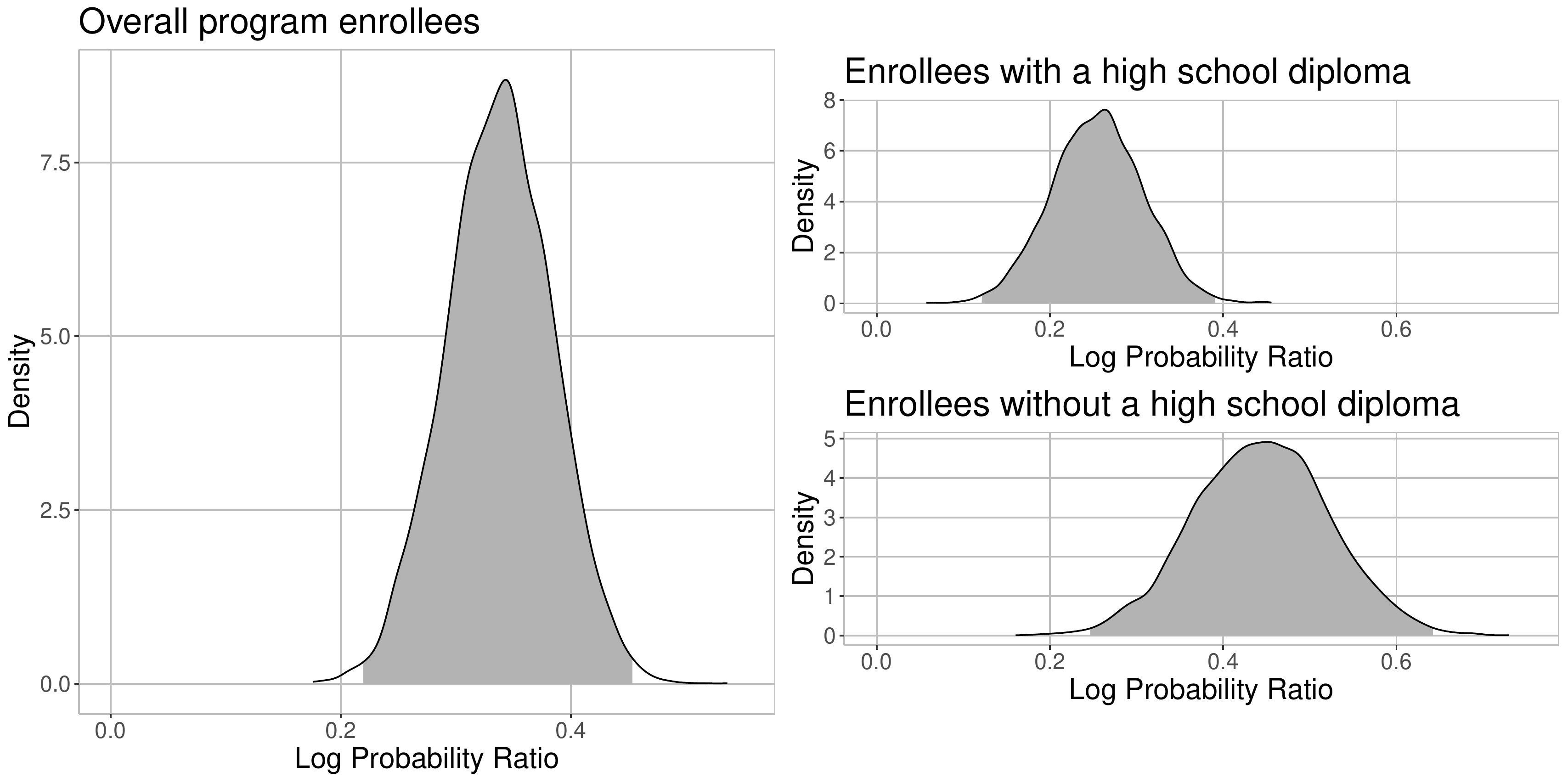}
\caption{Posterior distribution of the average log probability ratio, $\gamma_{avg}$, using the catalytic prior for the SWIM data. 
Left: the posterior distribution over all the enrollees; 
right: the posterior distributions over enrollees with a high school diploma and over the enrollees without a high school diploma, respectively. 
The shaded areas correspond to the 99\% credible intervals. 
}
\label{fig:SWIM-full}
\end{figure*}

We applied the catalytic prior to find the posterior distributions of $\bmbeta_{t}$ and $\bmbeta_{c}$, from which we can draw inference about various causal estimands. Here we consider the average log probability ratio, $\gamma_{avg}$, defined in \cref{eq:avg_gamma}.
\cref{fig:SWIM-full}, the left panel, shows the posterior distribution of $\gamma_{avg}$ over all the program enrollees. The 99\% credible interval of $\gamma_{avg}$ is given by $[0.219, 0.454]$, which is entirely on the positive real line, indicating that the SWIM program increases the enrollees' probability of employment. Our findings are consistent with those of \cite{friedlander1993saturation}, which found that the SWIM program accelerated employment. 
We can also study the causal effect on specific subgroups of the program enrollees. 
For example, we can delve into the effect on enrollees with and without a high school diploma. 
The right panel of \cref{fig:SWIM-full} shows two posterior distributions of $\gamma_{avg}$: one for those holding a high school diploma (comprising 56.1\% of the total enrollees) and another for those without (43.9\% of enrollees).
The means of the two posterior distributions are 0.255 and 0.444. 
Their respective 99\% credible intervals are given by $[0.121, 0.391]$ and 
$[0.246, 0.643]$. 
These results suggest that the SWIM program increases the probability of employment for both subgroups, but possibly more for those without a high school diploma. 

\subsection{Analysis on partial samples} \label{sec: SWIM partial samples}
We next illustrate how the catalytic prior performs in comparison to other methods by considering the hypothetical situation where only a part of the data is available to the researcher. 
We take a small random subsample of the data and apply different methods to analyze it. 
We then compare the results obtained from the competing methods to the benchmark, which we take as the posterior estimate based on the \textit{full} data using the flat prior, because the sample size of the full data is large enough ($N=3211$) for standard inference on the two logistic regressions (each with 12 unknown parameters) to yield stable results. Please note that under a flat prior, the posterior mode is the same as the MLE.

Method A works better than Method B on the small subsample, if the result from Method A is closer to the benchmark, which uses the full data, than the result obtained from Method B. We compare three methods: the flat prior (corresponding to MLE), the Catalytic prior, and the Cauchy prior, which was proposed in \citet{gelman2008weakly} specifically for logistic regression models
\footnote{When using the Cauchy prior for a logistic regression model, 
the nonbinary regressors are required to be rescaled to have mean 0 and standard deviation 0.5, and then independent Cauchy distributions are placed as priors on the coefficients. We follow the recommendation in \citet{gelman2008weakly} and use the default Cauchy distribution with center 0 and scale 2.5.}. 

In our experiment, we draw a random subsample of size $n$ ($n<N$) from the full data ($n/2$ units from the treatment group and $n/2$ units from the control group), and then fit the two logistic regressions based on this small training data set of size $n$ using three different priors: the catalytic prior, the Cauchy prior, and the flat prior.
For each choice of the causal estimand, we can compute the posterior mean of the causal estimand and compare it with the benchmark (which is the posterior mean given the full data using the flat prior). 
We focus on the average log probabilities ratio $\gamma_{avg}$ defined in \cref{eq:avg_gamma} over all enrollees in the subsample as well as $\gamma_{avg}$ over various subgroups, for example, enrollees with and without a high school diploma. We repeat the subsampling experiment independently 250 times for each $n\in \{100,  200, 400, 800, 1600\}$.

\cref{tab:swim-post-summary-mse-se} summarizes the posterior means of $\gamma_{avg}$ given by the three different methods based on subsamples of various sizes compared to the posterior mean of $\gamma_{avg}$ based on the full data using the flat prior (the benchmark).  
``MSE'' in the table refers to the mean squared error of the estimator based on subsamples against the benchmark, calculated from averaging over the 250 replications; ``SE'' refers to the standard error of the MSE, also calculated from the 250 replications. 
The smallest MSE in each case is in boldface. 
In addition to the subgroups according to the education level, we can consider other subgroups in our analysis, for example, subgroups based on age, gender, marital status, etc. \cref{tab:swim-post-summary-all} in the \hyperref[appn:swim summary]{Appendix} reports the analysis for eight subgroups. 

\cref{tab:swim-post-summary-mse-se} shows that uniformly for each subsample size $n$ and each subgroup, the posterior estimate given by the catalytic prior is the closest to the benchmark in MSE, followed by the estimate resulting from the Cauchy prior. 
When $n$ is as small as $200$, the flat prior does not provide a stable estimate, as indicated by its large MSE. As the subsample size $n$ increases, posterior estimates from these methods gradually converge to the benchmark, which is expected. 

\begin{table*}
    \centering
\begin{tabular}{ll|ccc}
\hline
& & \multicolumn{3}{c}{MSE (SE)} \\ 
Group & $n$  & Catalytic & Cauchy & Flat \\ 
\hline 
All & 100  & \textbf{0.141} (0.016) & 0.722 (0.109) & $> 50$ ($> 50$) \\
 & 200  & \textbf{0.060} (0.007) & 0.104 (0.012) & $> 50$ ($> 50$) \\
 & 400  & \textbf{0.021} (0.002) & 0.025 (0.002) & 0.027 (0.003) \\
 & 800  & \textbf{0.007} (0.001) & 0.008 (0.001) & 0.008 (0.001) \\
 & 1600  & \textbf{0.003} ($< 0.001$) & \textbf{0.003} ($< 0.001$) & \textbf{0.003} ($< 0.001$) \\
\hline 
With Highschool & 100  & \textbf{0.147} (0.023) & 0.654 (0.114) & $> 50$ ($> 50$) \\
Diploma & 200  & \textbf{0.060} (0.006) & 0.097 (0.010) & $> 50$ ($> 50$) \\
 & 400  & \textbf{0.028} (0.003) & 0.033 (0.004) & 0.035 (0.004) \\
 & 800  & \textbf{0.009} (0.001) & 0.010 (0.001) & 0.010 (0.001) \\
 & 1600  & \textbf{0.003} ($< 0.001$) & \textbf{0.003} ($< 0.001$) & \textbf{0.003} ($< 0.001$) \\
\hline 
Without Highschool & 100  & \textbf{0.286} (0.028) & 1.597 (0.288) & $> 50$ ($> 50$) \\
Diploma & 200  & \textbf{0.136} (0.019) & 0.247 (0.035) & $> 50$ ($> 50$) \\
 & 400  & \textbf{0.052} (0.005) & 0.066 (0.007) & 0.074 (0.007) \\
 & 800  & \textbf{0.018} (0.002) & 0.020 (0.002) & 0.021 (0.002) \\
 & 1600  & \textbf{0.006} (0.001) & 0.007 (0.001) & 0.007 (0.001) \\
\hline 
\end{tabular}
    \caption{Summary of the posterior means of $\gamma_{avg}$ based on subsamples of various sizes given by three different methods (columns) compared to the posterior mean of $\gamma_{avg}$ based on the full data. 
    The causal estimand here is the average log probability ratio $\gamma_{avg}$. We consider the causal estimate over the entire subsample (top panel) and over specific subgroups: enrollees with or without a high school diploma. In each case, boldface text denotes the smallest MSE among the three methods considered. }
    \label{tab:swim-post-summary-mse-se}
\end{table*}

As a further assessment of the different methods, we examine their prediction accuracy of the causal effects for units that have \textit{not} been observed. Specifically, we draw another subsample (named \textit{test set}) with size $n^\prime$ ($n^\prime/2$ units from the treatment group and $n^\prime/2$ units from the control group) from the full data that has no overlap with the first subsample (the \textit{training set}) and is thus not observed by the analyst. We then use the estimate $(\widehat{\bmbeta}_{t,n}, \widehat{\bmbeta}_{c,n})$ from each method based on the training set of size $n$ to compute the predictive 
treatment effect 
for each of the $n^\prime$ units in the test set 
$$
\widehat{\gamma}_{j,n} 
=\difflogit{\bmX_{j} \tp \widehat{\bmbeta}_{c,n}}{\bmX_{j} \tp \widehat{\bmbeta}_{t,n}}, \quad j \in \mbox{ test set},
$$
which is then compared to the benchmark prediction  
$$
\widehat{\gamma}_{j,N}^{BM} 
=\difflogit{\bmX_{j} \tp \widehat{\bmbeta}_{c,N}^{BM}}{\bmX_{j} \tp \widehat{\bmbeta}_{t,N}^{BM}}, \quad j \in \mbox{ test set}, 
$$
where $(\widehat{\bmbeta}_{t,N}^{BM},\widehat{\bmbeta}_{c,N}^{BM})$ is the estimate resulting from the flat prior using the full data (with sample size $N$). 
For our experiment, we chose $\allowbreak n\in \{100,  200, 400, 800, 1600\}$ and $n^\prime = 500$, and independently replicated the subsampling process 250 times.

For every training set and each of the three priors (the flat prior, the Catalytic prior, and the Cauchy prior), we computed the mean squared difference in predictive treatment effects (MSDPTE) for the test set:
$$
\mbox{MSDPTE}= \frac{1}{n^\prime} \underset{j\in \text{test set}}{\sum}  \left(\widehat{\gamma}_{j,n}- \widehat{\gamma}_{j,N}^{BM}  \right)^2,
$$
where we used the posterior means of $\bmbeta_{t}$ and $\bmbeta_{c}$ when calculating $\widehat{\gamma}_{j,n}$ and $\widehat{\gamma}_{j,N}^{BM}$. 
A smaller MSDPTE value indicates that the prediction, derived from a method using the training set of size $n$, is closer to the prediction based on the benchmark (the posterior estimate resulting from the flat prior based on the full data). 

\begin{figure}[!ht]
\centering
\includegraphics[width=0.7\textwidth]{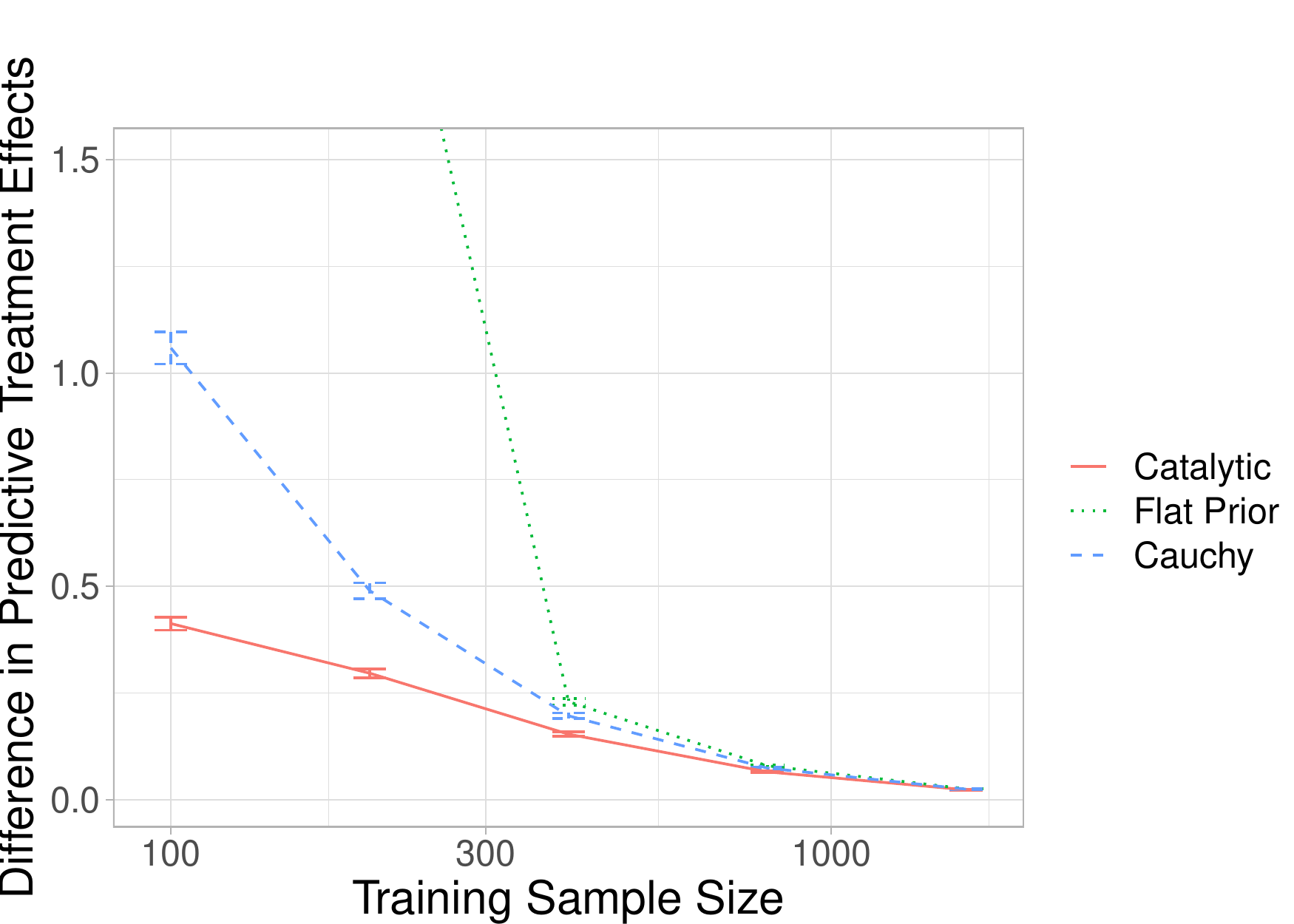}
\caption{Mean square difference in predictive treatment effects\  (MSDPTE) for the test set. The three curves (solid, dashed, dotted) represent the average MSDPTE of the posterior prediction resulting from, respectively, the catalytic prior, the Cauchy prior, and the flat prior over 250 independent replications. 
The error bars indicate one standard error from the 250 replications.
}
\label{fig:SWIM-ADPLOGATE}
\end{figure}

\cref{fig:SWIM-ADPLOGATE} shows the MSDPTE under the three different methods. 
Compared to the Cauchy prior, 
the predictions resulting from the catalytic prior are substantially more stable and closer to the benchmark prediction. 
When the training sample size $n$ is less than or equal to 200, the flat prior performs poorly relative to the other two methods. As $n$ increases, the predictions given by all three methods converge to the benchmark prediction. 

The numerical results in this example demonstrate the usefulness of the catalytic prior, especially when estimating a causal effect with a small sample that the standard method cannot fit reliably. 

\subsection{Alternative ways to generate synthetic data}\label{sec: more sources of synthetic data}

In this subsection, we examine alternative ways to generate synthetic data and the performances of the resulting posterior inferences. We begin by considering four alternative simpler models, $g$, that can be used to generate synthetic data, where each model contains only two parameters: the intercept term and the coefficient for a single covariate. The four respective covariates for the four simpler models are: 
\begin{enumerate}
    \item Having a high-school diploma (HSDip);
    \item Divorced or widowed (DivWidow); 
    \item Earnings in the previous year (EarnPre1y);
    \item Having positive earnings in the previous year (EmpPre1y).
\end{enumerate}
These four models, and the intercept-only model of the previous subsections, illustrate the hypothetical situation when a researcher might have multiple simpler models to contemplate. Here, in addition to the five individual simpler models, we can use a mixture of the five predictive distributions (with equal weights) to generate the synthetic data. This mixture generation reflects model averaging: given the uncertainty of the simpler models, instead of 
fixing on one particular choice of $g$, we consider ``averaging'' over the five models by mixing the synthetic data generated from each individual candidate model's predictive distribution.

\begin{figure}[!ht]
\centering
\includegraphics[width=0.7\textwidth]{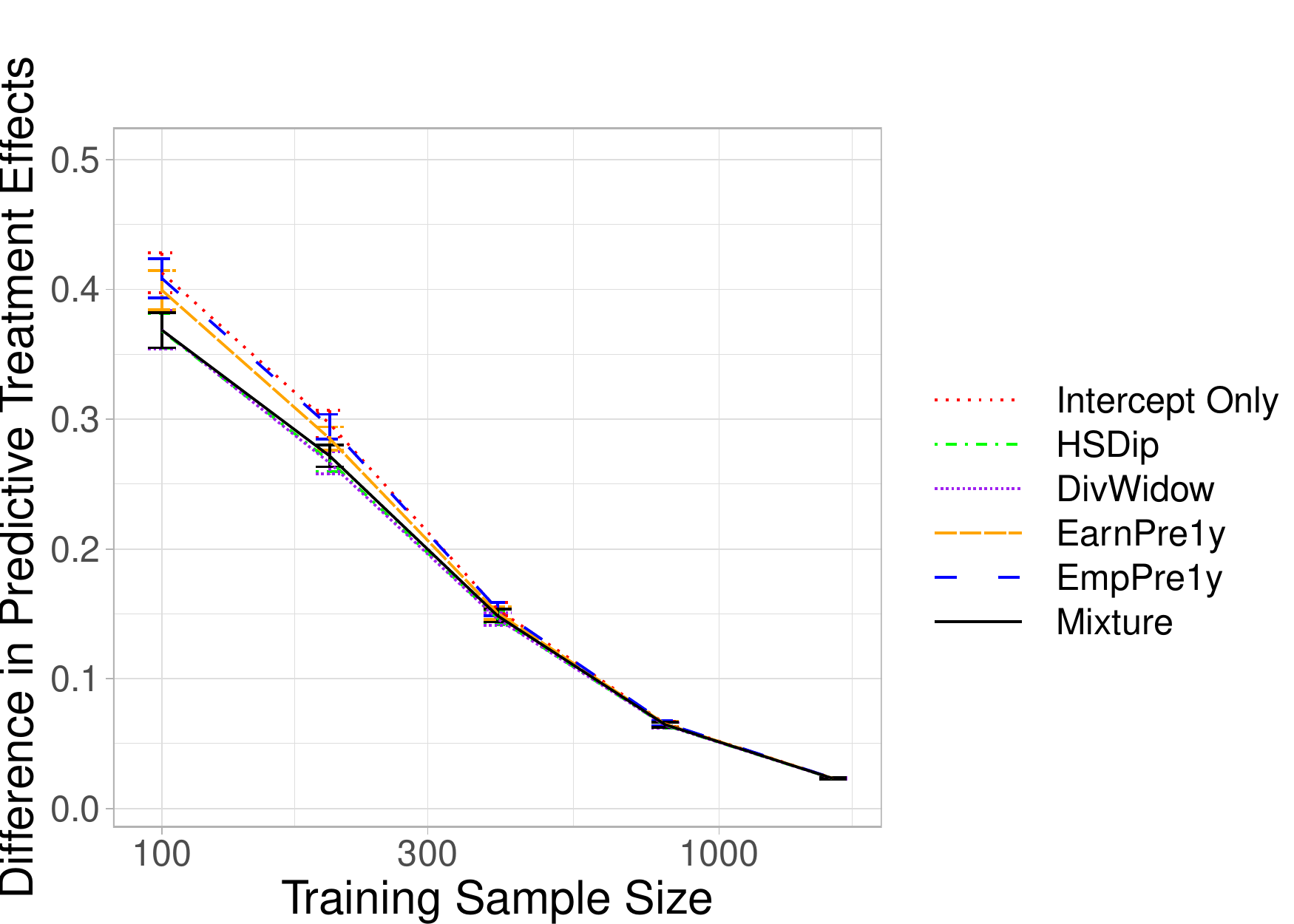}
\caption{Mean square difference in predictive treatment effects\ (MSDPTE) for the test set for various choices of synthetic data generation. 
Each curve represents the average MSDPTE of the posterior prediction resulting from the catalytic prior based on a particular way of generating the synthetic data (five individual models and their equal-weight mixture); the bottom three (almost overlapping) curves  correspond to the results by using the model with the covariate ``HSDip'', the model with the covariate ``DivWidow'', and the mixture data generation.
The error bars correspond to one standard error over 250 independent replications. 
}
\label{fig:SWIM-ADPLOGATE-g}
\end{figure}

Using the catalytic priors constructed from the synthetic data generated by the different ways (five individual distributions and the equal-weight mixture), we replicate the experiment from \cref{sec: SWIM partial samples} and calculate the mean squared difference in predictive treatment effects (MSDPTE) under each catalytic prior. \cref{fig:SWIM-ADPLOGATE-g}
plots the MSDPTE of the six cases. It is interesting to note that (i) the MSDPTEs of the other five cases are comparable to (and slightly better than) the MSDPTE based on the original intercept-only model, and (ii) the mixture synthetic data generation appears to offer the nearly smallest MSDPTE.
The results indicate that in our experiment on the SWIM dataset, varying the choice of the synthetic data generation can potentially improve the resulting inference, and that mixing the synthetic data generation (corresponding to the idea of model averaging) might lead to competitive catalytic priors.

\section{Connection with regularization methods in linear models}

Here we focus on linear models and study the connection between catalytic priors and several popular regularization methods. 
Both catalytic priors and common regularization methods pull the estimates of parameters towards something; 
in the former case, the shrinkage is towards the simpler model, whereas in the latter case, the shrinkage is towards zero. 
We first consider the connection between the catalytic prior approach and ridge regression (in \cref{sec:link-ridge}). 
The connection with other regularization methods is discussed after we view the distribution of the synthetic covariates as an entity that one can specify (\cref{sec:connect lasso,sec:elastic,sec:connect Lq,sec:connect group lasso}):
if we place some appropriate hyperpriors on the distribution of the synthetic covariates, the resulting posterior mode becomes identical to the LASSO \citep{tibshirani1996regression}, the elastic net \citep{zou2005regularization}, the $L_q$ penalized least-squares estimator \citep{frank1993statistical}, or the Group LASSO \citep{yuan2006model}. 

Throughout this section, the total weight parameter $\tau$ is considered a fixed positive number and we work on the population version of the catalytic prior as in \cref{eq: pop cata}. 
For simplicity, we assume that the covariates and the responses are centered at zero, and thus we do not include an intercept term in the linear model. 

\subsection{Connection with ridge regression}\label{sec:link-ridge}

Consider the linear regression model discussed in \cref{sec:example linear regression}: $\bmY = \bbX \bmbeta + \bmvarepsilon$, where the vector error term $\bmvarepsilon\sim  \calN( \bs{0}, \sigma^2 \mathbb{I}_n)$ with known $\sigma$, and the parameter to be estimated is $\bmbeta \in \bbR^p$. 
Suppose we specify a catalytic prior where the simpler model $g$ that generates the synthetic response is a sub-model of the linear regression model. 
Given the synthetic covariate vector $\vX^*$, the synthetic response is drawn from $Y^{*} | \boldsymbol{X}^{*} \sim\calN\left(\tilde{\boldsymbol{\beta}}_{0}^{\top} \boldsymbol{X}^{*}, \sigma^{2}\right)$, where  $\tilde{\boldsymbol{\beta}}_{0}$ is the fitted parameter of the simpler model $g$. 
Then under the population catalytic prior, the negative log posterior density is 
\begin{equation}
    \label{eq: loss func}
    \begin{aligned}
   &  \frac{1}{2 \sigma^2 }\norm{\bmY-\bbX\bmbeta}^2  \\ 
    & + \frac{1}{2\sigma^2} (\bmbeta-\tbmbeta_0)^\top \tau \EE{\bX^*(\bX^*)^\top} (\bmbeta-\tbmbeta_0),
    \end{aligned}
\end{equation}
and the posterior mode  is 
\begin{equation}
    \label{eq: LS cat hbeta}
   \begin{aligned}
    \hbmbeta = & \left(\bbX^\top \bbX + \tau \EE{\bX^*(\bX^*)^\top}\right)^{-1} \times \\
& \quad  \left(\bbX^\top \bmY + \tau \EE{\bX^*(\bX^*)^\top} \tbmbeta_0\right).
   \end{aligned}
\end{equation}

In the following special case, the estimate given by \cref{eq: LS cat hbeta} reduces to the well-known ridge regression estimate. 
Assume that the response is centered and that the covariates are both centered and standardized; when the synthetic covariates have no correlation and the simpler model is the constant model, then $\EE{\bX^*(\bX^*)^\top} = \mathbb{I}_p$ and $\tbmbeta_0 = 0$. 
Consequently, 
\cref{eq: LS cat hbeta,eq: loss func} match ridge regression with the objective function $\norm{\bmY-\bbX\bmbeta}^2 + \tau \norm{\bmbeta}^2$ and the point estimate of $\bmbeta$ 
\begin{equation*}
    \hbmbeta_{ridge} = \left(\bbX^\top \bbX + \tau \mathbb{I}_p\right)^{-1}
\bbX^\top \bmY.
\end{equation*}
In this special case, there is no correlation between different elements of the synthetic covariates $\bX^*$, the synthetic response $Y^*$ is drawn from $\calN(0,\sigma^2)$, and the catalytic prior pulls the estimate of $\bbeta$ towards the origin. For general cases, let
$\Delta := \tau \EE{\bX^*(\bX^*)^\top}$. Then \cref{eq: loss func} becomes 
\begin{equation*}
\norm{\bmY-\bbX\bmbeta}^2 + (\bmbeta-\tbmbeta_0)^\top \Delta (\bmbeta-\tbmbeta_0),
\end{equation*}
and \cref{eq: LS cat hbeta} becomes 
\begin{equation*}
\hbmbeta_{ridge} = \left(\bbX^\top \bbX + \Delta \right)^{-1}(\bbX^\top \bmY + \Delta \tbmbeta_0),
\end{equation*}
which are exactly the same as with generalized ridge regression \citep{hoerl1970ridge}.
The connection between ridge regression and the use of prior data has been discussed earlier in \cite{marquaridt1970generalized} and \cite{jackson1979use}.

\subsection{Connection with the  LASSO}\label{sec:connect lasso}

To demonstrate the connection between catalytic priors and the LASSO, 
first consider the distribution of the synthetic covariates. 
We will place a specific hyperprior on the scale of each synthetic covariate and show that the posterior mode resulting from the catalytic prior is closely related to the LASSO. 

Specifically, assume that the covariates are centered and standardized. We first draw an initial synthetic covariate vector  $\bX^{*,0}$ by independent sampling: let  $\bX^{*,0}=\left( X^{*,0}_{1}, \ldots, X^{*,0}_{p} \right)\tp $ where each component is independently resampled from the marginal distribution of the observed covariates. 
Each coordinate of $\bX^{*,0}$ is then scaled individually to give a synthetic covariate vector  
\begin{equation}
\label{eq: lasso synthetic data}
    \bX^*=\left( \frac{X^{*,0}_{1}}{\sqrt{s_1}}, \ldots, \frac{X^{*,0}_{p}}{ \sqrt{s_p}} \right)\tp,
\end{equation} 
where $s_1,\dots,s_{p}$ are positive hyperparameters. 
The conditional covariance matrix of the synthetic covariates, given $s_1,\dots,s_{p}$, is 
\begin{equation}
\label{eq: synthetic data cov}
    \bSigma_{X^*} = \bD_s \cdot  \bSigma_{0} \cdot \bD_s,
\end{equation}
where $\bSigma_{0}$ is the covariance matrix of $X^{*,0}$ and $\bD_s= \diag{1/\sqrt{s_1},\dots,1/\sqrt{s_{p}} } $.  

Because the original observed covariates are centered and standardized,
we have $\bSigma_{0} = \mathbb{I}_p$ and  $\bSigma_{X^*} = \diag{1/s_1,\dots,1/s_{p}}$. 
We assign independent exponential distributions on $s_j$'s as hyperprior distributions so that each has a density proportional to $\exp(-\frac{\lambda}{\tau} s_j)$. The exponential priors prevent the $s_j$'s from being too large, thereby ensuring that the synthetic covariates are reasonably dispersed.
Given this choice of the hyperprior, the negative log joint posterior density can be simplified as
\begin{equation}
   \label{eq:opt-design-scale}
 \frac{1}{2 \sigma^2 }  \|\vY-\mX\bmbeta\|^{2}
 +\frac{\tau}{2 \sigma^2 } \sum_{j=1}^{p} \frac{1}{s_j}(\beta_j-\tilde{\beta}_{0,j})^{2}+\frac{\lambda}{\tau} \sum_{j=1}^{p} s_j .
\end{equation}
The posterior mode is found by minimizing \cref{eq:opt-design-scale} over $\bmbeta$ and $s_{j}$'s jointly. 
By first optimizing over $s_{j}$'s, the optimization reduces to
\[
\min_{\bmbeta} \left\{ \frac{1}{2}  \|\vY-\mX\bmbeta\|^{2}
+ \sqrt{2\lambda\sigma^2} \sum_{j=1}^{p} |\beta_j-\tilde{\beta}_{0,j}| \right\}.
\label{eq: cat to lasso}
\] 
If the simpler model is the constant model, then $\tilde{\beta}_{0,j}=0$ 
and the optimization is the same as that of the LASSO \citep{tibshirani1996regression}. 

\subsection{Connection with the elastic net} \label{sec:elastic}

The elastic net  incorporates both the $L_2$ and $L_1$ regularization \citep{zou2005regularization}. 
We can establish a connection between the catalytic prior and the elastic net by using the idea of generating mixture synthetic data as discussed in Sections~\ref{sec:cata spec} and \ref{sec: more sources of synthetic data}. 

Specifically, suppose we generate synthetic data with equal probability from two different sources: 
\begin{enumerate}
    \item  $\vX^*$ with a fixed covariance matrix $\bSigma_{0}$, and $Y^{*} | \boldsymbol{X}^{*} \sim  \calN\left(\bar{\boldsymbol{\beta}}_{0}^{\top} \boldsymbol{X}^{*}, \sigma^{2}\right)$ as in Section~\ref{sec:link-ridge};
    \item $\vX^*$ with a scaled covariance matrix    $\bSigma_{X^*} = \bD_s \times  \bSigma_{0} \times \bD_s$,
where the scaling matrix is $\bD_s= \diag{1/\sqrt{s_1},\dots,1/\sqrt{s_{p}} }$, and $Y^{*} | \boldsymbol{X}^{*} \sim \calN\left(\tilde{\boldsymbol{\beta}}_{0}^{\top} \boldsymbol{X}^{*}, \sigma^{2}\right)$ as in Section~\ref{sec:connect lasso}. 
\end{enumerate}
For simplicity, we consider the case where  $\bSigma_{0}=\mathbb{I}_p$. 
As in Section~\ref{sec:connect lasso}, we assign independent exponential distributions on $s_j$'s as hyperprior distributions. 
Under this specification of the population catalytic prior, \cref{eq: pop cata}, the negative log posterior density is 
\begin{equation*}
    \label{eq: loss enet}
    \begin{aligned}
   &  \frac{1}{2 \sigma^2 }\norm{\bmY-\bbX\bmbeta}^2  + \frac{\tau}{4\sigma^2} \|\bmbeta-\bar{\bmbeta}_0\|^2 \\ 
&  +\frac{\tau}{4 \sigma^2 } \sum_{j=1}^{p} \frac{1}{s_j}(\beta_j-\tilde{\beta}_{0,j})^{2}+\frac{\lambda}{\tau} \sum_{j=1}^{p} s_j .
    \end{aligned}
\end{equation*}
Now consider the joint posterior mode of $\bmbeta$ and $s_{j}$. By minimizing over $s_j$'s first, the posterior mode of $\bmbeta$ satisfies the following optimization equation: 
\[
\begin{aligned}
\min_{\bmbeta} \left\{ \frac{1}{2}  \|\vY-\mX\bmbeta\|^{2}
 \right. & + \frac{\tau}{4\sigma^2} \|\bmbeta-\bar{\bmbeta}_0\|^2  \\ 
  & \left. + \sqrt{\lambda\alpha \sigma^2} \sum_{j=1}^{p} |\beta_j-\tilde{\beta}_{0,j}| \right\},
\label{eq: cat to enet}
\end{aligned}
\] 
which corresponds to the optimization expression in the elastic net.

\subsection{Connection with $L_q$ regularization}
\label{sec:connect Lq}
We can generalize the discussion in \cref{sec:connect lasso} by setting a more general hyperprior on each $s_j$ with a density function proportional to $\exp(-\frac{\lambda}{\tau} s_j^{r})$, where $r$ is a positive tuning constant.  \cref{eq:opt-design-scale} now becomes
    \begin{equation*}
         \frac{1}{2 \sigma^2 }  \|\vY-\vX\bmbeta\|^{2} +\frac{\tau}{2 \sigma^2 } \sum_{j=1}^{p} \frac{1}{s_j}(\beta_j-\tilde{\beta}_{0,j})^{2}+\frac{\lambda}{\tau} \sum_{j=1}^{p} s_{j}^{r} .
    \end{equation*}
\noindent By optimizing over $s_j$'s, the problem reduces to:
    \begin{equation*}
    \begin{aligned}
          \min_{\bmbeta} & \left\{   \frac{1}{r+1} \|\vY-\mX\bmbeta\|^{2}+ \right. \\
&  \left.  \quad        \left(\frac{\tau^{r-1}2\lambda\sigma^2}{r^r}\right)^{\frac{1}{r+1}}
        \sum_{j=1}^{p} |\beta_j-\tilde{\beta}_{0,j}|^{\frac{2r}{r+1}} \right\},
    \end{aligned}
    \end{equation*}
    which is the same as $L_q$ regularization with $q = \frac{2r}{r+1}$ \citep{frank1993statistical}. By changing the value of $r$, we can obtain any $L_q$ penalization for $q \in (0,2)$.

\subsection{Connection with the Group LASSO}
\label{sec:connect group lasso}
In many cases, the covariates can be naturally grouped a priori. For instance, a group is formed by the dummy variable representation of a categorical covariate. Assume that the indices $1, 2, \dots, p$ are partitioned into $g$ groups, denoted as $G_i$ ($i=1,2,\dots,g$), and that the covariates within the same group share the same scale factor instead of $p$ possibly different ones as in \cref{sec:connect lasso}.
Without loss of generality, we can  reorder the indices to ensure $(1, 2, \dots, p)=(G_1, \ldots, G_g)$. 
If we independently sample each group of an initial synthetic covariate vector  $\bX^{*,0}$, then there is no correlation between covariates in different groups, and  $\bSigma_{0}$ (the covariance matrix of $\bX^{*,0}$) can be written in block diagonal form: 

   \begin{equation*}
    \bSigma_{0} = 
    \begin{bmatrix}
         \bSigma_{G_1} &0            &\cdots  & 0\\
         0           & \bSigma_{G_2} &\cdots & 0\\
         \vdots      & \vdots      &\ddots & 0 \\
         0           & \cdots      &0      & \bSigma_{G_g} \\
    \end{bmatrix}
\end{equation*}
Analogous to \cref{eq: lasso synthetic data}, which assigns a separate scale factor to each covariate, we now assign a separate scale factor to each covariate group of $\bX^{*,0}$ to obtain $\bX^{*}$: 
\begin{equation*}
\label{eq: group synthetic data}
    \bX^*=\left( X^{*,0}_{G_1}/ \sqrt{\bar{s}_1}, \ldots, X^{*,0}_{G_g} / \sqrt{\bar{s}_g} \right)\tp
\end{equation*} 
with a covariance matrix  
    \begin{equation*}
        \bSigma_{X*} = 
            \diag{\bar{s}_1^{-1}  \bSigma_{G_1} ,  ,\dots, \bar{s}_g^{-1}  \bSigma_{G_g} }.       \end{equation*}
    
We consider independent exponential hyperpriors given by $\bar{s}_i \sim \exp(-\frac{\lambda}{\tau} \bar{s}_i)$. Following the derivation of \cref{eq: cat to lasso}, the posterior mode of $\bmbeta$ is the solution to
    \[\label{eq: Group Lasso penalty}
\min_{\bmbeta} 
\left\{ 
    \frac{1}{2}  \|\vY-\mX\bmbeta\|^{2}+
    \sqrt{2\lambda\sigma^2}   \sum_{i=1}^g \norm{\beta_{G_i} - \tilde{\beta}_{0,G_i}}_{\bSigma_{G_i}} 
\right\},
\] 
where $\beta_{G_i}$ is the sub-vector of $\bmbeta$ corresponding to $G_i$ and
$\norm{v}_{\bSigma_{G_i}} = \sqrt{v^{\top}\bSigma_{G_i}v}$.
 The optimization in \cref{eq: Group Lasso penalty} is almost the same as the Group LASSO \citep{yuan2006model}; here the norm of the sub-vector $\beta_{G_i}$ is determined by a positive definite matrix $\bSigma_{G_i}$, whereas in the ordinary Group LASSO, the norm is the Euclidean norm. Nevertheless, the solution of \cref{eq: Group Lasso penalty} preserves the desirable property of the Group LASSO:  the estimate either selects or excludes an entire group of variables.

\section{Conclusion} 
This article discusses a method to construct catalytic prior distributions, especially apposite when the data are not sufficient to stably fit a complex target model. Such a prior distribution augments the observed data with synthetic data that are sampled from the predictive distribution of a simpler model fit to the observed data. 
The simpler model can be (but is not limited to) a sub-model of the complex model. 
Because the catalytic prior construction uses synthetic data (rather than focusing on the model parameters), it is easier to interpret the information included by a catalytic prior. 
The catalytic prior construction is adaptive to many models because the expression of a catalytic prior is the weighted likelihood of synthetic data.
In particular, when numerous models need fitting, making it impossible to manually specify a reasonable prior for each, the catalytic prior provides an automated, systematic, and consistent specification. 
The computation procedure is generally no more difficult than generating the synthetic data and fitting the complex model using standard software. 
Therefore, the class of catalytic priors is also computationally attractive for complex Bayesian models.

The usefulness of catalytic priors was illustrated here in a real-world example from labor economics.
In this example, the resulting inference captures important aspects of the data set and exhibited competitive or superior estimation and prediction accuracy, compared to inferences based on other commonly used priors.

We also examined the connections of catalytic priors with various regularization methods for linear models. 
Catalytic priors effectively shrink the estimated parameters of the complex target model towards a simpler model whereas the other regularization methods typically shrink toward zero. Furthermore, catalytic priors are constructed using synthetic data whereas the others are constructed using penalty functions, which are often less intuitive than synthetic data.

The construction of the catalytic prior is data-dependent.
Such a data dependent construction can be viewed as an empirical Bayes method in which the hyper-parameters (such as the ones in the simpler model) are estimated from the data. This idea of data-dependent specification can be traced back at least to the classical Box-Cox transformation \citep{Box:1964ev}, where the power transformations of variables are determined by the data.

Catalytic priors are known to enjoy several attractive properties. 
When applied to generalized linear models, \cite{huang2020catalytic} proved that catalytic priors are proper under mild conditions. The properness of catalytic priors implies that they are suited for use in Bayesian model comparison, such as the calculation of Bayes factors. They also demonstrated in simulations that the resulting inferences have frequentist operating characteristics that are competitive with other proposed methods. 
Catalytic priors may thus be applicable to a wide range of complex Bayesian models and could be useful in routine Bayesian analysis.

\section{Acknowledgments}
We thank the reviewers and editors for their constructive comments that  improved the presentation of this manuscript. 
% Grant
D. Huang was partially supported by NUS Start-up Grant A-0004824-00-0 and Singapore Ministry of Education AcRF Tier 1 Grant A-8000466-00-00. 
D. B. Rubin currently also holds faculty appointments at Tsinghua University (in Beijing, China) and Temple University (in Philadelphia). 
The SWIM data used in this paper are derived from data files made available to researchers by MDRC. The authors are solely responsible for how the data set has been used or interpreted.

\bibliography{ref}

\clearpage
\newgeometry{top=1in,left=1in,textwidth=6.8in,textheight=9in}

\begin{appendix}
\section{Summary of Subgroup Posterior Analysis for the SWIM Example}\label{appn:swim summary}

Section \ref{sec: SWIM partial samples} compares the three different prior distributions. This Appendix expands the comparison to additional subgroups, including the subgroups based on education level, age, and marital status (ever married, divorced or widowed). \cref{tab:swim-post-summary-all} displays the analysis on eight different subgroups.

\setcounter{table}{0}
\renewcommand{\thetable}{A.\arabic{table}}

\begin{center}
\begin{longtable}{ll|ccc}
    \caption{Summary of the posterior means of $\gamma_{avg}$ based on subsamples of various sizes given by three different methods compared to the posterior mean of $\gamma_{avg}$ based on the full data. 
    The causal estimand here is the average log probability ratio $\gamma_{avg}$. We consider the causal estimand over the whole subsample, as well as over several subgroups. 
    hsdip+ (or $-$): enrollees who had (or lacked) a high school diploma; 
    age $>$ (or $\leq$) 35: enrollees who are older than 35 (or not); 
    newmar+ (or $-$): enrollees who never (or ever) married; 
    divwid+ (or $-$): enrollees who were divorced or widowed (or not).
    In each case, boldface corresponds to the smallest MSE among the three methods considered.
    }  \label{tab:swim-post-summary-all}
\\
%%%
\hline
& & \multicolumn{3}{c}{MSE (SE)} \\ 
Group & $n$  & Catalytic & Cauchy & Flat \\ 
\hline 
 \endfirsthead
%%%
\multicolumn{5}{l}{Table A.1 (Continued).} \\ 
\hline
& & \multicolumn{3}{c}{MSE (SE)} \\ 
Group & $n$  & Catalytic & Cauchy & Flat \\ 
\hline 
\endhead 
All & 100  & \textbf{0.141} (0.016) & 0.722 (0.109) & $> 50$ ($> 50$) \\
 & 200  & \textbf{0.060} (0.007) & 0.104 (0.012) & $> 50$ ($> 50$) \\
 & 400  & \textbf{0.021} (0.002) & 0.025 (0.002) & 0.027 (0.003) \\
 & 800  & \textbf{0.007} (0.001) & 0.008 (0.001) & 0.008 (0.001) \\
 & 1600  & \textbf{0.003} ($< 0.001$) & \textbf{0.003} ($< 0.001$) & \textbf{0.003} ($< 0.001$) \\
\hline hsdip$+$ & 100  & \textbf{0.147} (0.023) & 0.654 (0.114) & $> 50$ ($> 50$) \\
 & 200  & \textbf{0.060} (0.006) & 0.097 (0.010) & $> 50$ ($> 50$) \\
 & 400  & \textbf{0.028} (0.003) & 0.033 (0.004) & 0.035 (0.004) \\
 & 800  & \textbf{0.009} (0.001) & 0.010 (0.001) & 0.010 (0.001) \\
 & 1600  & \textbf{0.003} ($< 0.001$) & \textbf{0.003} ($< 0.001$) & \textbf{0.003} ($< 0.001$) \\
\hline hsdip$-$ & 100  & \textbf{0.286} (0.028) & 1.597 (0.288) & $> 50$ ($> 50$) \\
 & 200  & \textbf{0.136} (0.019) & 0.247 (0.035) & $> 50$ ($> 50$) \\
 & 400  & \textbf{0.052} (0.005) & 0.066 (0.007) & 0.074 (0.007) \\
 & 800  & \textbf{0.018} (0.002) & 0.020 (0.002) & 0.021 (0.002) \\
 & 1600  & \textbf{0.006} (0.001) & 0.007 (0.001) & 0.007 (0.001) \\
\hline age$>$35 & 100  & \textbf{0.221} (0.023) & 0.989 (0.135) & $> 50$ ($> 50$) \\
 & 200  & \textbf{0.104} (0.011) & 0.189 (0.022) & $> 50$ ($> 50$) \\
 & 400  & \textbf{0.038} (0.004) & 0.048 (0.005) & 0.053 (0.005) \\
 & 800  & \textbf{0.015} (0.002) & 0.017 (0.002) & 0.018 (0.002) \\
 & 1600  & \textbf{0.005} ($< 0.001$) & \textbf{0.005} (0.001) & \textbf{0.005} (0.001) \\
\hline age$\leq$35 & 100  & \textbf{0.183} (0.021) & 0.965 (0.170) & $> 50$ ($> 50$) \\
 & 200  & \textbf{0.073} (0.008) & 0.120 (0.013) & $> 50$ ($> 50$) \\
 & 400  & \textbf{0.030} (0.003) & 0.036 (0.004) & 0.039 (0.004) \\
 & 800  & \textbf{0.012} (0.001) & \textbf{0.012} (0.001) & 0.013 (0.001) \\
 & 1600  & \textbf{0.004} ($< 0.001$) & \textbf{0.004} ($< 0.001$) & \textbf{0.004} ($< 0.001$) \\
\hline 
newmar$+$ & 100  & \textbf{0.302} (0.038) & 1.611 (0.425) & $> 50$ ($> 50$) \\
 & 200  & \textbf{0.123} (0.015) & 0.204 (0.027) & $> 50$ ($> 50$) \\
 & 400  & \textbf{0.052} (0.005) & 0.062 (0.006) & 0.068 (0.007) \\
 & 800  & \textbf{0.020} (0.002) & 0.021 (0.002) & 0.022 (0.002) \\
 & 1600  & \textbf{0.006} (0.001) & 0.007 (0.001) & 0.007 (0.001) \\
\hline 
newmar$-$ & 100  & \textbf{0.159} (0.015) & 0.753 (0.116) & $> 50$ ($> 50$) \\
 & 200  & \textbf{0.074} (0.008) & 0.129 (0.015) & $> 50$ ($> 50$) \\
 & 400  & \textbf{0.029} (0.003) & 0.035 (0.003) & 0.039 (0.004) \\
 & 800  & \textbf{0.010} (0.001) & 0.011 (0.001) & 0.012 (0.001) \\
 & 1600  & \textbf{0.003} ($< 0.001$) & \textbf{0.003} ($< 0.001$) & \textbf{0.003} ($< 0.001$) \\
\hline 
divwid$+$ & 100  & \textbf{0.217} (0.024) & 0.886 (0.123) & $> 50$ ($> 50$) \\
 & 200  & \textbf{0.115} (0.012) & 0.197 (0.021) & $> 50$ ($> 50$) \\
 & 400  & \textbf{0.046} (0.004) & 0.057 (0.005) & 0.063 (0.006) \\
 & 800  & \textbf{0.018} (0.002) & 0.020 (0.002) & 0.021 (0.002) \\
 & 1600  & \textbf{0.005} ($< 0.001$) & 0.006 (0.001) & 0.006 (0.001) \\
\hline 
divwid$-$ & 100  & \textbf{0.197} (0.031) & 1.054 (0.188) & $> 50$ ($> 50$) \\
 & 200  & \textbf{0.072} (0.008) & 0.130 (0.016) & $> 50$ ($> 50$) \\
 & 400  & \textbf{0.028} (0.003) & 0.034 (0.003) & 0.037 (0.004) \\
 & 800  & \textbf{0.011} (0.001) & \textbf{0.011} (0.001) & 0.012 (0.001) \\
 & 1600  & \textbf{0.004} ($< 0.001$) & \textbf{0.004} ($< 0.001$) & \textbf{0.004} ($< 0.001$) \\
\hline 
\end{longtable}
\end{center}

\end{appendix}

\end{document}